\RequirePackage{fix-cm}
\documentclass{svjour3}                     
\smartqed  
\usepackage{graphicx}
\usepackage{multirow}
\usepackage{subcaption}
\usepackage{tabularx}
\usepackage{xparse}
\usepackage{mathtools}
\newsavebox{\mybox}
\NewDocumentEnvironment{strange}{}
    {%
        \lrbox\mybox
        \tabularx{0.8\textwidth}{|l|X|}%
    }
    {%
        \endtabularx
        \endlrbox
    }

\usepackage{upquote}
\usepackage{booktabs}

\usepackage{times}
\journalname{Int J CARS}

\begin{document}

\title{The Effect of Video Playback Speed on Perception of Technical Skill in Robotic Surgery}
 


\author{Jason D. Kelly         \and
		Nicholas Heller \and
		Ashley Petersen Ph.D. \and
	     Thomas S. Lendvay M.D. \and
          Timothy M. Kowalewski Ph.D.
}


\institute{Jason Kelly \at
              Department of Mechanical Engineering, University of Minnesota, Minneapolis, MN \\
              \email{kell1917@umn.edu}           
		\and
		N. Heller \at
		Department of Computer Science \& Engineering, University of Minnesota, Minneapolis, MN
		\and 
		A. Petersen \at
		Division of Biostatistics, University of Minnesota, Minneapolis, MN 
		\and
		T. Lendvay \at
		Department of Urology, University of Washington, Seattle, WA
           \and
           T. Kowalewski \at
              Department of Mechanical Engineering, University of Minnesota, Minneapolis, MN
}

\date{Received: date / Accepted: date}

\maketitle

\begin{abstract}
\textit{Purpose:} Finding effective methods to evaluate surgeon technical skill has proven a complex problem to solve computationally. Previous research has shown that obtaining non-expert crowd evaluations of surgical performances concords with the gold standard of expert surgeon review, and that faster playback speed increases ratings for videos of higher-skilled surgeons in laparoscopic simulation \cite{CSATS_Paper}, \cite{Me}. The aim of this research is to extend this investigation to real surgeries that use non-expert crowd evaluations. We address two questions (1) whether crowds award more favorable ratings to videos shown at increased playback speeds, and (2) if crowd evaluations of the first minute of a surgical procedure differ from crowd evaluations of the entire performance.  \\ 
\textit{Methods:} A set of 56 videos of practicing (non-novice) surgeons including robotic prostatectomy, hysterectomy, and partial nephrectomy (for 28 ``expert" surgeons, and 28 who are ``proficient"), were used to evaluate the perceived technical skill of the surgeons at each video playback speed used (0.4x, 1.2x, 2.0x, 2.8x,  and 3.6x) for the first minute of the previously rated performance, using the Global Evaluative Assessment of Robotic Skills (GEARS) assessment criteria. Each video was subsequently rated at 1x speed to obtain objective ratings for the first minute of the surgical procedure.\\
\textit{Results:} Crowds on average did rate videos higher as playback speed was increased. This effect was observed for both proficient and expert surgeons. Each increase in the playback speed by 0.8x was associated with, on average, a 0.16-point increase in the GEARS score for expert surgeons and a 0.27-point increase in GEARS score for proficient surgeons, with both groups being perceived as obtaining relatively equal skill at the fastest playback speed. It was also found that 22 out of the 56 surgeons were perceived to be significantly different in skill when just viewing the first minute of performance, with 11 of the 28 surgeons in both skill categories being rated as belonging to the opposing category. \\
\textit{Conclusion:} The observed increase in skill ratings with video playback speed replicates findings for laparoscopic experts in \cite{Me}, and extends to the context of real robotic surgeries. The change in perceived technical skill due to increased playback speed for experts and proficient surgeons suggests that crowds do seem biased in rating surgeons as more highly skilled when they appear quicker, even if speeds seems unrealistic. Furthermore, the large differences in skill labels when comparing the first minute of surgery to the entire 15 minute video warrants further investigation into how much perceived skill ratings vary in time (sub-task level) vs. summative metrics (task level). 
\keywords{Crowd Sourcing \and Video Playback  \and Surgical Technical Skill \and Speed Perception \and Bias}
\end{abstract}

\section{Introduction}
\label{intro}

A third of all deaths in the United States are caused by medical errors, and surgical errors are one of the largest contributors to this \cite{ThirdLeadingDeath}. Technical surgical skill is directly related to patient outcomes \cite{Birkmeyer}, but it remains a difficult computational task to correctly classify surgeons into skill levels with a compelling level of accuracy, i.e. never misclassifying an `obvious novice' as an `obvious expert' and vice versa - the MAC Criterion \cite{MAC}. The de facto gold standard for evaluating technical skill is video evaluation by an expert surgeon using Likert-scale assessment metrics, in which evaluators submit ratings on an anchored scale of 1-5. Using crowds of \textit{non-expert} evaluators are a surprisingly accurate way to inexpensively and rapidly obtain skill level ratings for videos of surgical performances, with a pass/fail rating that matches 100\% of pass/fail ratings by expert surgeons \cite{CSATS}. The fact remains, however, that humans can be biased in their thinking, and subjective metrics of rating performances can lead to results we would not expect from computational models of evaluation. 

\begin{figure}[h!]
\centering
	\includegraphics[width = 0.6\textwidth]{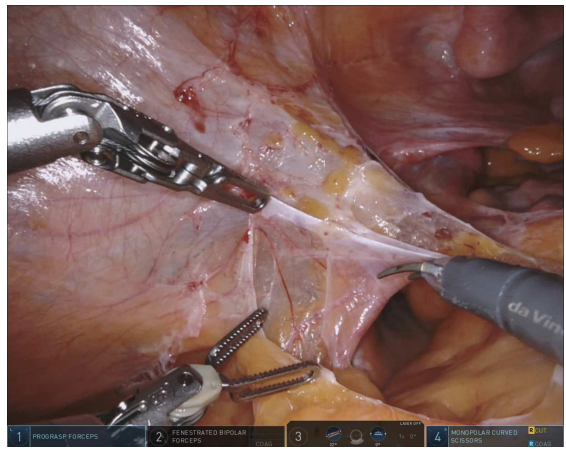}
	\caption{Frame from a video collected for this study, recorded from a daVinci surgical robot, Intuitive (Sunnyvale, CA).}
	\label{fig:RSR}
\end{figure}

A popular laparoscopic surgical skill assessment metric is the Global Evaluative Assessment of Robotic Skills (GEARS), which is the most common assessment tool for robotic surgery skills \cite{GEARS_Paper1}. The subdomains in this metric include: bimanual dexterity, efficiency, depth perception, force sensitivity and robotic control. Task time is not a direct metric used to evaluate the technical skill of laparoscopic surgeons with tools like GEARS, however time for task completion is often seen as one of the most predictive objective forms of evaluating technical skill \cite{BeyondTaskTime}, \cite{Hung}. However, there must be ways of objectively evaluating skill between multiple performances which were completed in the same time span.

\begin{figure}[ht]
\centering

	\begin{subfigure}{.3\textwidth}
		\includegraphics[width=\linewidth]{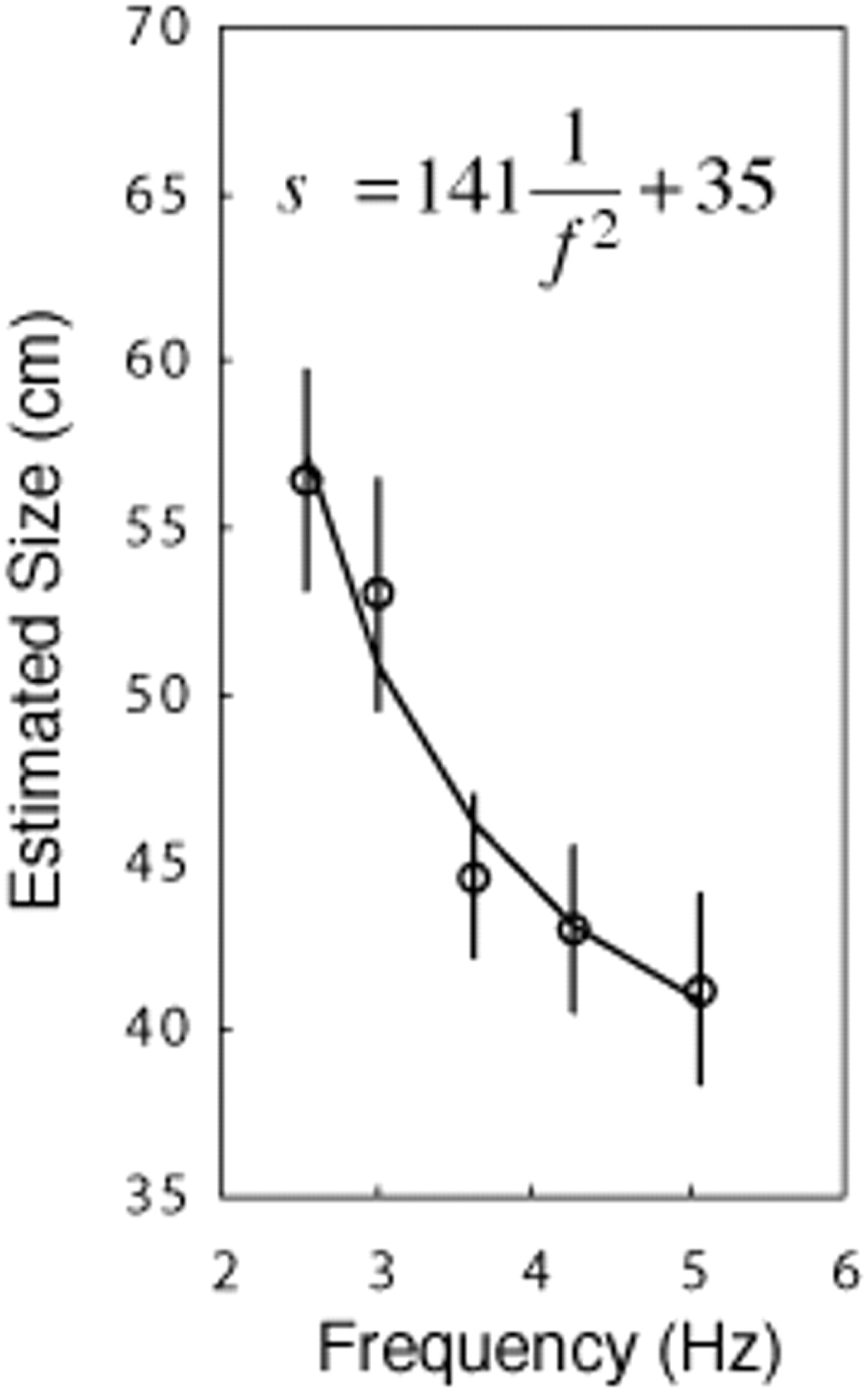}
	\end{subfigure}
	\begin{subfigure}{.6\textwidth}
		\includegraphics[width=\linewidth]{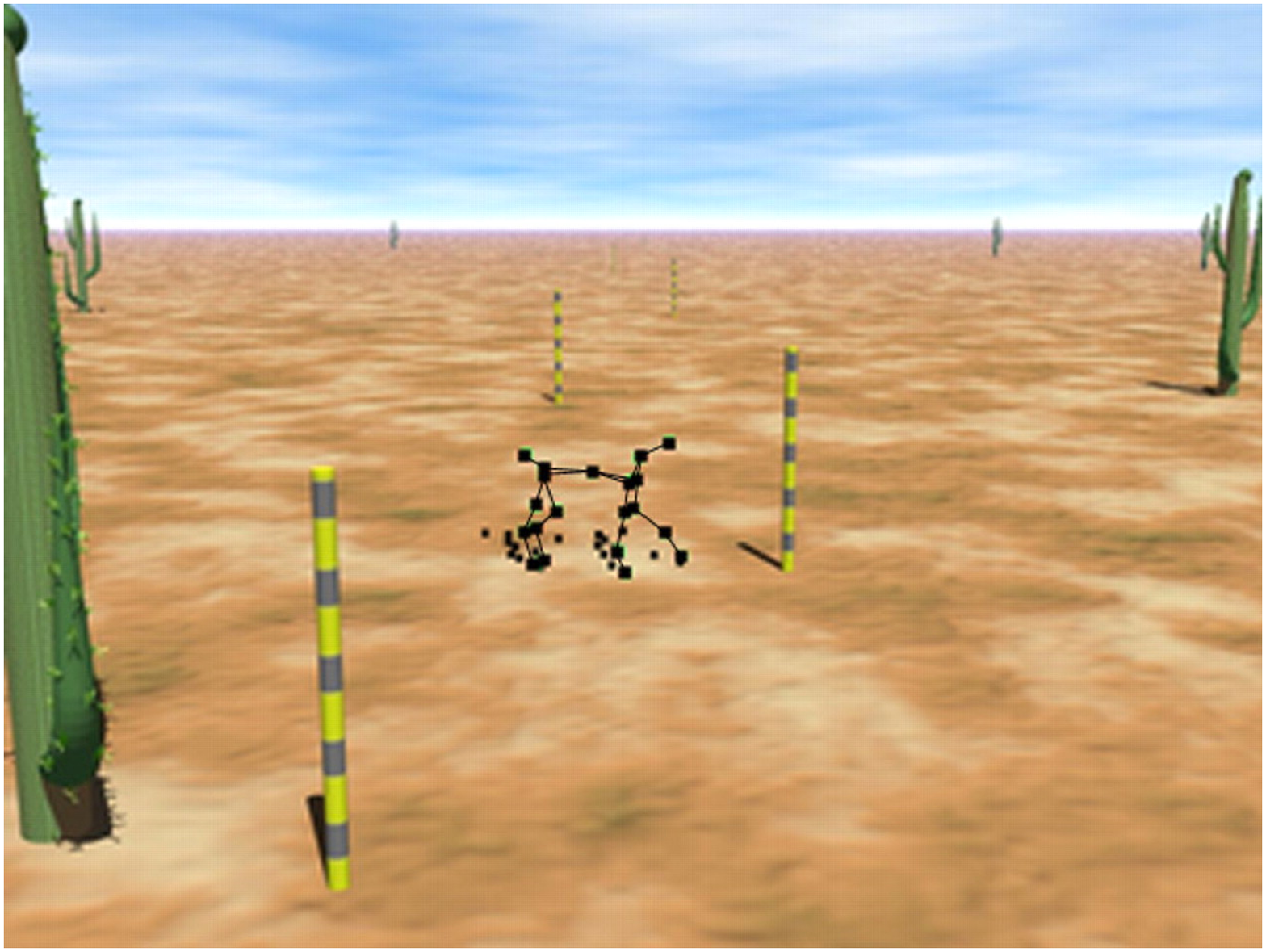}
	\end{subfigure}
	
	\caption{ Illustration of research results (left) which found the frequency of gait in simulated animals (right) affects the estimated size \cite{Jokische}.}
	\label{fig:size_Perception}
	
\end{figure}

\subsection{Changes in Perception due to Speed}
\label{BiologicalMotionPerception}

Research in biological motion processing has observed a relationship between the gait frequency and estimated size of animals and other objects which replicate the joints of the human body. It was found that when a computer simulation was used to artificially modify the gait speed of these objects to unnaturally high speeds, study participants perceived the size of the moving objects as changing. As the gait frequency was increased, the estimated size of the object enlarged, shown in Fig. \ref{fig:size_Perception}, \cite{Jokische}, \cite{altered_PLW}. The results from this study may suggest a link between how humans evaluate biological motion and the speed of a movement. It is possible this phenomenon extends to other areas of evaluation, such as technical skill in surgery. 

\subsection{Technical Skill Ranges}
\label{SkillRanges}
Previous work has shown that crowds are able to discriminate the levels of novice and expert surgeons and their scores change as playback speed changes - even into motions appearing artificially sped up \cite{Me}. This was a limited study using a small dataset of laparoscopic training exercises from the Basic Laparoscopic Urologic Skills dataset \cite{Initial_BLUS_curriculum}, \cite{Locations_BLUS}, with extremes of skill levels (e.g.  ``obvious novice" medical students with no suturing experience). To test whether this phenomenon relates to actual surgical procedures, a larger dataset must be used which incorporates videos of real surgical footage and more practical ranges of surgeon skill. However, surgeons which are allowed to safely operate on patients can't ethically be ``obvious" novice surgeons, according to the MAC Criterion \cite{MAC}. This means the surgeons in a dataset of real surgical footage will comprise a smaller range of technical skill levels, with all surgeons being more comparably rated. This necessitates the need to further separate experts into two groups: proficient and expert surgeons. Proficient surgeons as stated in \cite{proficient} are defined as a surgeon who is well advanced in any branch of knowledge or skill. We will adopt this as our term for surgeons who are in the bottom quintile of scores from our surgical videos, as they are still well above average skilled surgeons. 

Our motivation in this work is to investigate whether, by increasing video playback speed in real robotic surgery, surgeons will be perceived by crowds as more skilled when they appear to be moving faster, \textit{and} how these effects vary from expert and proficient surgeons. We will do this by evaluating the ability of non-expert crowd workers to discriminate proficient surgeons from experts. Finally we seek to examine the effect of video duration on ratings following \cite{Anna_NSeconds}. 

\begin{table}[h!]
\caption{Likert-scale technical skill perception questionnaire, with five of the six domains of the GEARS assessment tool, excluding autonomy.}
\label{tab:GEARS_Questions}
\begin{strange}\hline
Score & \textbf{Depth Perception}  \\ \hline \hline
(1) & Constantly overshoots target, wide swings, slow to correct    \\ \hline
(2) &   \\ \hline
(3) & Some overshooting or missing of target, but quick to correct \\ \hline
(4) &  \\ \hline
(5) & Accurately directs instruments in the correct plane to target \\ \hline
\end{strange}
\begin{center}
\usebox\mybox
\end{center}
\begin{strange}\hline
Score & \textbf{Bimanual Dexterity}\\ \hline \hline
(1) & Uses only one hand, ignores non-dominant hand, poor\\
    & coordination\\ \hline
(2) &   \\ \hline
(3) & Uses both hands, but does not optimize interaction \\
    & between hands \\ \hline
(4) & \\ \hline
(5) & Expertly uses both hands in a complementary way to\\
    & provide optimal exposure \\ \hline
\end{strange}
\begin{center}
\usebox\mybox
\end{center}
\begin{strange} \hline
Score & \textbf{Efficiency} \\ \hline \hline
(1) & Inefficient efforts; many uncertain movements; \\ 
    &   constantly changing focus or persisting without progress \\ \hline
(2) &   \\ \hline
(3) & Slow, but planned movements are reasonably organized  \\ \hline
(4) &   \\ \hline
(5) & Confident, efficient and safe conduct, maintains focus \\
    & on task, fluid progression \\ \hline
\end{strange}
\begin{center}
\usebox\mybox
\end{center}
\begin{strange} \hline
Score & \textbf{Force Sensitivity}\\ \hline \hline
(1) & Rough moves, tears tissue, injures nearby structures, poor control,\\  
  & frequent suture breakage \\ \hline
(2) &   \\ \hline
(3) & Handles tissue reasonably well, minor trauma to adjacent \\
    & tissue, rare suture breakage\\ \hline
(4) &   \\ \hline
(5) & Applies appropriate tension, negligible injury to adjacent \\
    & structures, no suture breakage \\ \hline
\end{strange}
\begin{center}
\usebox\mybox
\end{center}
\begin{strange} \hline
Score & \textbf{Robotic Control}\\ \hline \hline
(1) & Consistently does not optimize view, hand position, or repeated collisions even with guidance\\ \hline
(2) &   \\ \hline
(3) & View is sometimes not optimal. Occasionally needs to relocate arms. \\
    & Occasional collisions and obstruction of assistant.\\ \hline
(4) &   \\ \hline
(5) & Controls camera and hand position optimally and independently. \\
    & Minimal collisions or obstruction of assistant. \\ \hline
\end{strange}
\begin{center}
\usebox\mybox
\end{center}
\end{table}

\section{Methods}
\label{Methods}
\subsection{Dataset}
\label{Dataset}

This study used a novel Robotic Surgery Readiness (RSR) Study dataset, which consists of 343 videos of live robotic surgeries, with matching kinematic data also recorded, though unused here. These surgeries were performed by attending surgeons and trainees in urology, gynecology, and general surgery at the University of Washington Medical Center and the Puget Sound Veterans Administration. Each video was manually edited to include roughly the first 15 minutes of surgical activity performed by the surgeon. There are a wide variety of surgical procedures recorded from this dataset including: prostatectomy, cystectomy, hysterectomy, partial nephrectomy, and sacrocolpopexy. An image of a frame from one of these surgeries is shown in Fig. \ref{fig:RSR}. A GEARS score for each performance was also obtained from crowd evaluation. 

The range of scores for the RSR videos was fairly small, with most lying between 20-22 out of 25. In an effort to get the largest possible range of skill from this dataset, 28 performances from the top quintile of scored performances and 28 from the bottom quintile of performances were used and given labels of `expert' and `proficient', respectively, keeping in mind that almost all of the performances would have objectively been considered expert-like. For semantic analysis, the first minute of surgical activity was extracted from each 12-15 minute video for analysis by crowds. 

Amazon Mechanical Turk was the crowd-sourcing platform used for this study, in which each non-expert crowd worker was paid an average of \$0.40 to watch and evaluate a video. A web domain was created for which Turkers would be redirected to, where they submitted a consent form and were asked questions about videos. Two different kinds of experiments were conducted: \textit{technical skill perception at different playback speeds} and \textit{sub-task level skill labeling}. 

\subsection{Technical Skills Perception at Different Playback Speeds}
\label{Experiment1}

Technical skill perception was measured by surveying non-expert crowds to give each video performance a GEARS score by rating each of the 5 subdomains shown in Table \ref{tab:GEARS_Questions}. Forty ``turkers" were recruited independently for each video, in which each video at each playback speed was independently submitted to the website in order to avoid a grouping bias. Videos were altered to play at 0.4x, 1.2x, 2.0x, 2.8x, and 3.6x, edited using FFmpeg \cite{FFMPEG}, in which frames were either taken out or added in order to create the resulting playback speed. The score from each of five subdomains (Depth Perception, Bimanual Dexterity, Efficiency, Force Sensitivity, and Robotic Control) were summed to create a cumulative score for each performance in the range of 5-25.  

A linear mixed effects model was used to analyze the significance of the various speeds to the evaluations received. The significance of the time spent on reviewing a video and the labels given to the surgeons were also examined. It was hypothesized that each Mechanical Turker on the site should be assumed to have a different slope, to match the difference with which they relatively evaluate different videos. The evalutor ID given to each rater was assigned a random effect. In addition, fixed effects for both the speed at which the video was played and the amount of time spent on the video by the evaluator were controlled for in the model. The mixed model was compared with a null model, which did not include the fixed effects, using ANOVA hypothesis testing and comparing the relative information criteria and correlation between the scores and the various parameters. All data aggregation and statistics were calculated in Python 3.6 \cite{Python} and R \cite{R}.

\subsection{Sub-Task Level Skill Labeling}
\label{Experiment1}

To learn if the first minute of surgical activity displayed the same level of technical skill as the entire 15 minute video and should have the same label, the first minute of each video at 1x (normal playback speed) was submitted to crowds for review. The mean of these new scores were then calculated, and the 50th percentile was used as the cut-off point to signify in this dataset of 56 videos, whether a surgeon would be labeled as a proficient or expert level surgeon. The differences in these labels were then analyzed further and visualized, to find if the previously obtained scores for the entire 15 minute video possessed significantly different skill levels in the first minute, such that the label of the corresponding surgeon was different than what we had originally obtained. 


\section{Results}

\begin{table}[!h]
\caption{Results from the linear mixed effects model testing speed and time spent on evaluation}
\label{LME_Res}
\centering
\begin{tabular}{@{}lrrrrr@{}} \toprule
\textbf{Fixed Effects} & \textbf{Estimate} & \textbf{Standard Error} & \textbf{df}  & \textbf{t value} & \textbf{Pr ( \textgreater{} $\vert$ t $\vert$)} \\ \cmidrule(l){1-6}
\multicolumn{6}{ l }{\textbf{Initial linear mixed effects model (BIC = 52465.36)}}  \\ \cmidrule(l){1-6}
Speed   & 0.046 & 0.013 & 13360 & 3.47 & 5.31e-04  \\ 
Elapsed Time   & -0.0023 &  3.47e-4 & 13720  & -6.57 & 5.18e-11 \\ \cmidrule(l){1-6}
\multicolumn{6}{ l }{Final linear mixed effects model (BIC = 52410.60)}  \\ \bottomrule
\end{tabular}
\label{tab:LME_Res}
\end{table}

\begin{table}[]
\caption{The types of robotic surgery performed and demographics of the surgeons from the 56 videos used.}
\label{tab:Surgery_Type}
\centering
\begin{tabular}{@{}cc@{}} \toprule
\textbf{Surgery Type} & \textbf{N} \\ \cmidrule(l){1-2}
Prostatectomy & 17 \\ 
Hysterectomy & 11 \\
Nephrectomy & 5  \\
Cystectomy & 4 \\
Other & 19 \\ \bottomrule
\end{tabular}
\quad\quad\quad
\begin{tabular}{@{}cc@{}} \toprule
\textbf{Demographic} & \textbf{N} \\ \cmidrule(l){1-2}
Male & 17 \\ 
Female & 11 \\
Mean Age & 46.70 \\
Mean Yrs. Experience & 12.85 \\ \bottomrule
\end{tabular}
\end{table}

\subsection{Data Demographics}

Table \ref{tab:Surgery_Type} summarizes the types of robotic surgery and the demographic data from the $N=56$ videos (and surgeons) used in this work. 

\subsection{Technical Skill Perception}

\begin{figure}[h!]
  \centering
  \hspace{-0.8cm}
  \includegraphics[width=0.99\textwidth]{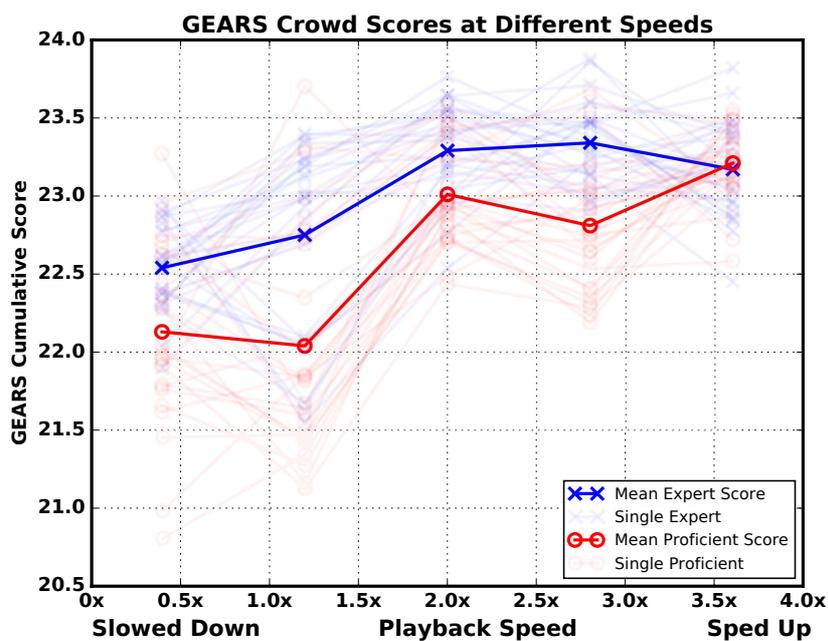}
\caption{All mean crowd evaluations (bold lines) from each proficient and expert surgeon at various video playback speeds. Each single surgeons video (semi-transparent colors) indicates ratings from N = 40 turkers.}
\label{fig:All_Scores}       
\end{figure}

\begin{figure}[h!]

	\centering
	
    \begin{subfigure}[t]{0.49\textwidth}
    
        \raisebox{-\height}{\includegraphics[width=\textwidth]{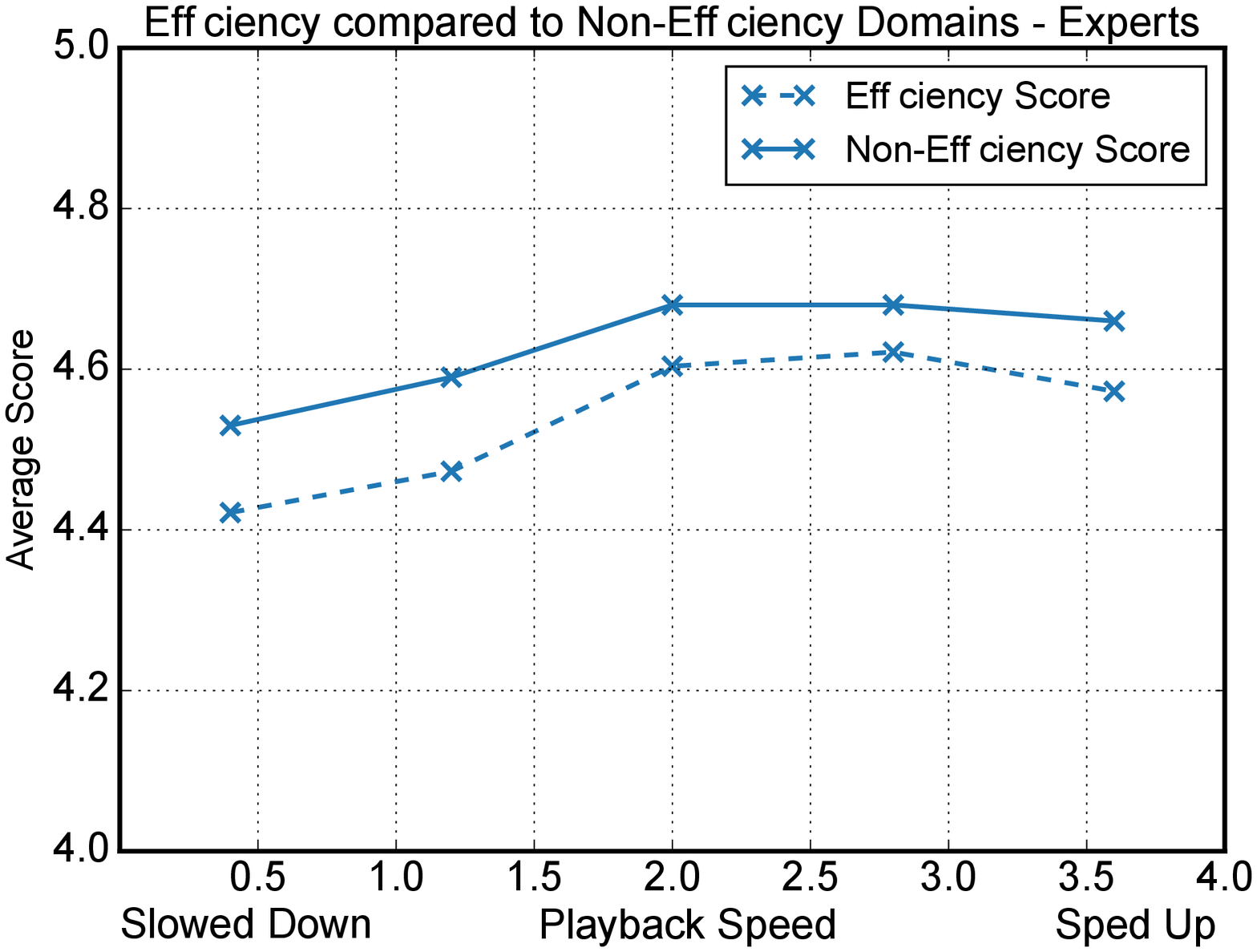}}
        
    \end{subfigure}
    \begin{subfigure}[t]{0.49\textwidth}
        \raisebox{-\height}{\includegraphics[width=\textwidth]{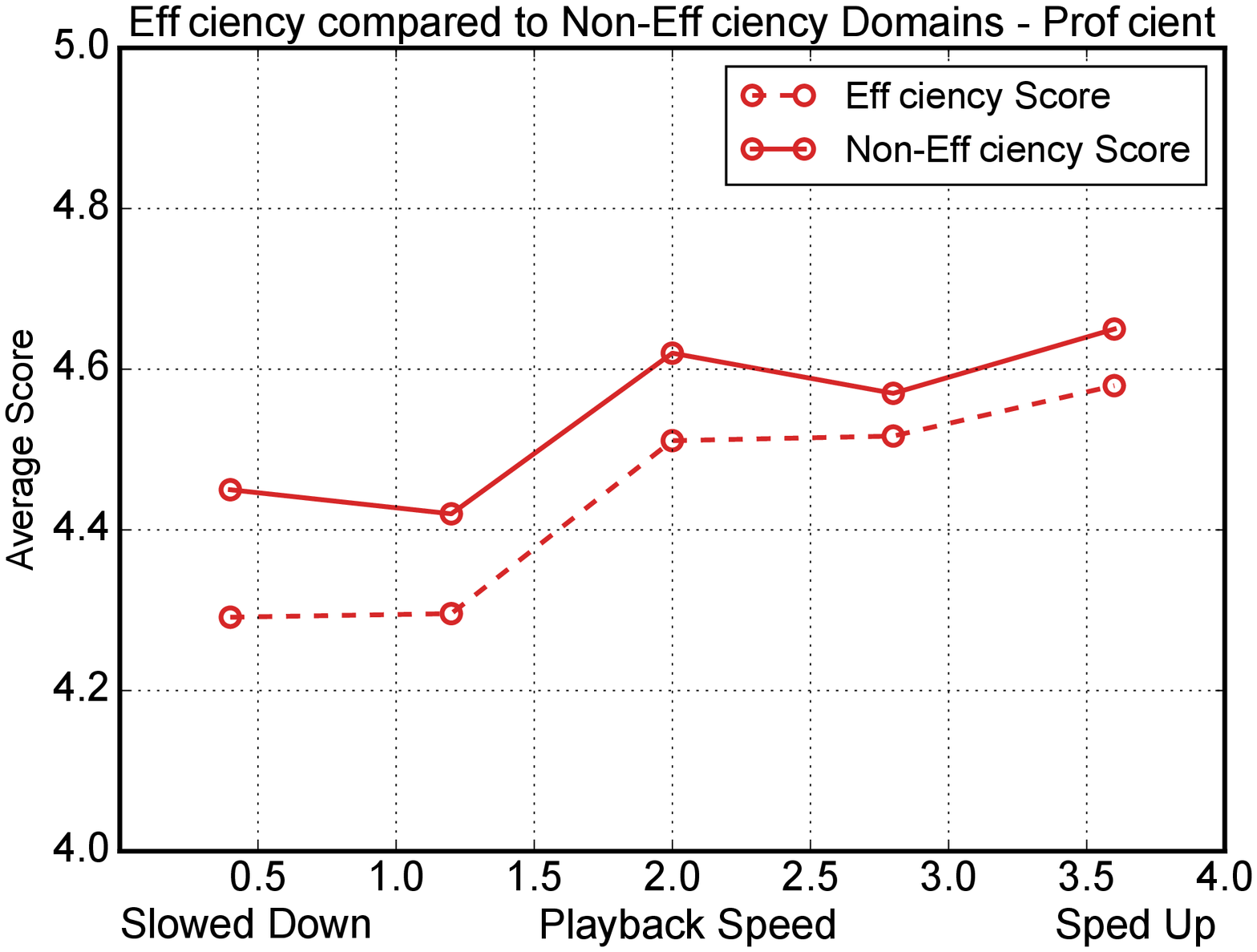}}
    \end{subfigure}
	\caption{The efficiency subdomain as compared to the mean of the other GEARS subdomains, for expert and proficient surgeons.}
	\label{fig:Stacked_Area}
	
\end{figure}
The mean of the GEARS score for each group and video are shown in Fig. \ref{fig:All_Scores}. For expert videos, each increase in the playback speed by 0.8x was associated with, on average, a 0.16-point increase in the GEARS score (95\% CI: 0.10-0.22 point increase; p $<$ 0.05). On average these scores appear to increase within a sublevel of the playback speeds around 0.4x to 2.0x, and then level out at all remaining playback speeds. For proficient surgeon videos, each increase in the playback speed by 0.8x was associated with, on average, a 0.27-point increase in the GEARS score (95\% CI: 0.19-0.35 point increase; p $<$ 0.05). Thus, while both experts and proficient surgeons experienced increased perceived technical skill as the playback speed was increased, the mean score obtained at the fastest video playback speed reached an almost equal skill level for both proficient and expert level performances. Fig. \ref{fig:Stacked_Area} shows the increase in the efficiency subdomain as well as the mean of the other four domains to visualize whether efficiency (seen as the most related to speed) is the only increasing domain. As shown, no major difference is apparent between the two types of domains.

\subsection{Sub-Task Level Labeling}

\begin{figure}[h!]
  \centering
  \hspace{-0.8cm}
  \includegraphics[width=0.99\textwidth]{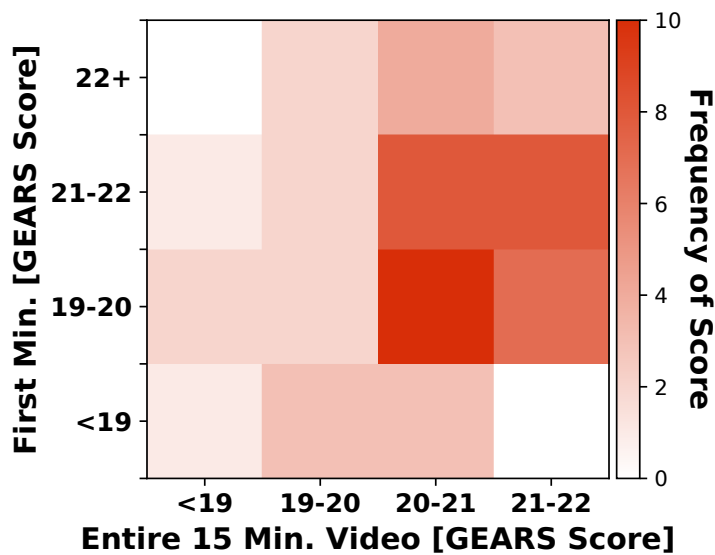}
\caption{GEARS scores given to the entire 15 minute video of a performance, compared to only the first minute of the same video.}
\label{fig:Conf_Matrix}       
\end{figure}

Figure \ref{fig:Conf_Matrix} illustrates the difference in GEARS scores by non-expert crowd workers, when viewing the first minute of the video compared to the entire video. As shown, there is a noticeable difference in the scores given at these two levels. Most surgeons, when viewed for the entire 15 minutes, are rated slightly higher than when only the first minute of surgical activity is performed. Viewing just the label given to the performance (proficient or expert), by analyzing whether the video was above or below the median score in the group of 56 videos, a total of 11 previously labeled proficient level surgeons, and 11 previously labeled expert surgeons switched the label they were originally given, when only the first minute of surgical activity was evaluated.

\section{Conclusion}

The results from the technical skill perception study give support to our initial hypothesis that increasing the video playback speed would increase the ratings of experienced surgeons. Now this evidence extends to real surgeries using robotic surgical performances. Surprisingly, however, we discovered that for sub-groups of expert level surgeons ( proficient and expert ) the increase in score happens at a quicker rate for proficient surgeons than for expert level surgeons. Increasing the playback speed of a slightly less than expert-level surgeon tends to ``wash out" the minor mistakes they have made in the performance, effectively making the two groups appear more equally skilled at higher playback speeds. Additionally from analyzing the `Efficiency' subdomain from the GEARS assessment, it appears that - surprisingly - crowds are also biased to give higher ratings in domains which are not associated with speed. 


The results from the sub-task level skill labeling experiment show us that it may be necessary to have video evaluated in smaller duration segments, due to the notable disagreement with the skill level assigned to the entire video vs. just using the first minute. This lends support to the notion that a surgeon's technical skill may fluctuate in time (on the order of minutes) throughout a surgical procedure. This warrants further study of non-constant skill mapped to a single summative rating, as this would induce substantial label noise of computational skill evaluation using machine learning. 

We conclude that increasing the video playback speed of performances of practicing surgeons in typical robotic surgeries results in increased scores as reported by non-expert crowds. This effect is surprisingly uniform across GEARS subdomains, even those which should be unaffected by speed. We further conclude that more studies should be done to investigate variance in time of sub-task level videos of surgical procedures, as the technical skill of a surgeon may fluctuate on the order of minutes or less during the procedure.

\section*{Compliance With Ethical Standards}
	
	\noindent\textbf{Conflicts of Interest}
	The authors declare that they have no conflict of interest.
	
	\noindent\textbf{Ethical Standard}
	All procedures performed in studies involving
	human participants were in accordance with the ethical standards of
	the institutional and/or national research committee and with the 1964
	Helsinki declaration and its later amendments or comparable ethical
	standards.
	
	\noindent\textbf{Funding}
	This work was supported, in part, by the Office of the Assistant Secretary of Defense for Health Affairs under Award No. W81XWH-15-2-0030, the National Science Foundation CAREER grant under Award No. 1847610, as well as the National Institutes of Health’s National Center for Advancing Translational Sciences, grant UL1TR002494. Opinions, interpretations, conclusions, and recommendations are those of the authors and are not necessarily endorsed by the Department of Defense, the National Science Foundation, or the National Institutes of Healths's National Center for Advancing Translational Sciences.

	\noindent\textbf{Informed Consent} Informed consent was obtained from all individual participants included in the study.

\bibliographystyle{spbasic}      


\end{document}